\documentclass[12pt]{article}
\usepackage{amsmath,amssymb,bm,amsthm}
\usepackage{hyperref}

\newcommand{\ket}[1]{| #1 \rangle}
\newcommand{\bra}[1]{\langle #1 |}
\newcommand{\Eq}[1]{Eq.~(\ref{#1})}
\newcommand{\mi}{i}
\newcommand{\me}{e}
\newcommand{\cc}[1]{{\mkern1mu\overline{\mkern-1mu #1\mkern-1mu}\mkern1mu}}
\newcommand{\brkt}[2]{\langle #1 | #2 \rangle} 

\newcommand{\ul}[1]{\underline{\smash{#1}\vphantom{0_0}}} 
\newcommand{\wb}[2]{\phi^#1_#2}
\newcommand{\kwb}[2]{\ket{\wb{#1}{#2}}}
\newcommand{\Id}{\mathbf 1}
\newcommand{\Wt}{\mathcal{W}}
\newcommand{\Wit}{\widetilde\Wt} 
\newcommand{\HWit}{\Wit_H}
\newcommand{\HZWit}{\Wit_{2H}}

\newcommand{\Sec}[1]{Section~\ref{Sec:#1}}
\newcommand{\Bas}{\mathcal{B}}

\title{Penrose Dodecahedron, Witting Configuration and Quantum Entanglement}
\date{}

\author{Alexander Yu.\ Vlasov}

\begin{document}
	
\sloppy

\maketitle

\begin{abstract}
A model with two entangled spin-3/2 particles based on geometry of dodecahedron 
was suggested by Roger Penrose for formulation of analogue of Bell theorem 
`without probabilities.' The model was later reformulated using so-called
Witting configuration with 40 rays in 4D Hilbert space. However, such 
reformulation needs for some subtleties related with entanglement of 
two such configurations essential for consideration of non-locality and some other
questions. Two entangled systems with quantum states described by Witting configurations
are discussed in presented work. Duplication of points with respect to vertices of dodecahedron 
produces rather significant increase with number of symmetries in $25920/60=432$ times. 
Quantum circuits model is a natural language for description of operations 
with different states and measurements of such systems.
\end{abstract}

\section{Introduction}
\label{Sec:Intro}

Specific configuration of 40 rays based on geometry of dodecahedron 
was suggested by Roger Penrose in \cite{Pdod,ZP} for demonstration of non-probabilistic 
version of Bell theorem. The construction was later popularized
in his book {\em Shadows of the Mind}  \cite{shadows}.

The {\em Majorana map} \cite{Ma32} between $n$-dimensional 
complex space and set with $n{-}1$ points on Riemann sphere
was used in suggested approach. 
Thus, three points can be used for description of state of spin-3/2 particle
in 4D Hilbert space. 

Positions of some points in the triple may coincide and the order is not prescribed, 
{\em e.g.}, in suggested approach for any vertex $V$ of dodecahedron a triple mapped 
into one of 20 `explicit' rays  was defined as $\{V,V,-V\}$, where $-V$ is opposite vertex. 
The less straightforward construction of yet another 20 `implicit' rays is obtained
as complements to full bases of some triples of orthogonal `explicit' rays. 

Symmetries of whole set with  40 rays in 4D Hilbert spaces considered as 40-point configuration
in complex projective space $\mathbb{CP}^3$ was also discussed in \cite{ZP}. 
It was mentioned an existence of ``transitive group of 25920 unitary projective
transformations of $\mathbb{CP}^3$ sending the configuration of 40 points 
into itself'' \cite{ZP}. The group is denoted further as $\Wt$.

The model was analyzed later by P. K. Aravind 
{\em et al} \cite{MA,WA} and it was found, that all 40 rays
can be represented in alternative way without Majorana map
using so-called Witting configuration. However, entanglement 
of two such configurations essential for discussions about 
non-locality was not considered.

The presented paper is devoted to consideration of some basic
properties of Witting configuration necessary for work with
two entangled systems. Any such system can be represented as
pair of qubits or `ququart'. Some useful properties and 
symmetries of single Witting configuration are recollected in \Sec{Wit}.
Entanglement of two such configurations is discussed in \Sec{Ent}.
Most analytic calculations are performed using GAP software \cite{GAP}.

\section{Witting configuration and polytope}
\label{Sec:Wit}

\subsection{Complex Witting polytope}
\label{Sec:poly}

Witting polytope was introduced by Coxeter \cite{CoxReg,Cox84}.
It has $240$ vertexes in complex space   
\begin{equation}\label{points240}
\begin{split}
 (0,\pm\omega^\mu,\mp\omega^\nu,\pm\omega^\lambda),\quad
 & (\mp\omega^\mu,0,\pm\omega^\nu,\pm\omega^\lambda), \\
 & (\pm\omega^\mu,\mp\omega^\nu,0,\pm\omega^\lambda),\quad
 (\mp\omega^\mu,\mp\omega^\nu,\mp\omega^\lambda,0), \\
 (\pm\mi\omega^\lambda \sqrt{3},0,0,0),\qquad
 & (0,\pm\mi\omega^\lambda \sqrt{3},0,0), \\
 & (0,0,\pm\mi\omega^\lambda \sqrt{3},0),\qquad
 (0,0,0,\pm\mi\omega^\lambda \sqrt{3}),
\end{split} 
\end{equation}
where $\omega = (-1+\mi \sqrt{3})/2$, $\omega^3=1$ and
$\lambda, \mu, \nu \in \{0,1,2\}$ are independent numbers. 
The polytope is named after configuration of 40 points 
in projective spaces suggested by Alexander Witting \cite{Wit1887}.
The {\em Witting configuration} with 40 rays in 4D Hilbert space
is defined up to phases $\pm\omega^\mu$, $\mu=0,\ldots,2$ and it is more appropriate 
for discussion about quantum states.

After normalization of \Eq{points240} there are $40$ rays described by unit vectors
\begin{subequations}\label{rays40}
\begin{equation}
\begin{split}\label{rays40a}
 \frac{1}{\sqrt{3}}(0,1,-\omega^\mu,\omega^\nu),\quad
 &\frac{1}{\sqrt{3}}(1,0,-\omega^\mu,-\omega^\nu), \\
 &\frac{1}{\sqrt{3}}(1,-\omega^\mu,0,\omega^\nu),\quad
 \frac{1}{\sqrt{3}}(1,\omega^\mu,\omega^\nu,0),
\end{split} 
\end{equation}
\begin{equation}\label{rays40b}
 (1,0,0,0),\qquad (0,1,0,0),\qquad
 (0,0,1,0),\qquad (0,0,0,1)
\end{equation}
\end{subequations}
with $\mu, \nu \in \{0,1,2\}$.
Each vector in \Eq{rays40} corresponds to six points of complex Witting polytope
in \Eq{points240} due to `phase multipliers'. In comparison with \cite{CoxReg}
the \Eq{rays40} use slightly different notation with normalization on 
{\em the first} nonzero coordinate. It coincides with convention used
in \cite{WA}.

Possible indexing for the states is shown in Table~\ref{tab40rays}.
Here $\kwb{l}0=\ket{l}$, $l=0,\dots,3$ is a `computational' basis \Eq{rays40b}
and the normalization $1/\sqrt{3}$ from \Eq{rays40a} 
for 36 states  $\kwb{l}{n}$ with $n \neq 0$ is omitted in Table~\ref{tab40rays} 
due to typographical and some other reasons. A correspondence of such indexing
with implicit and explicit rays and vertices of dodecahedron can be found in \cite{WA},
but it is rather excessive for presented work.

\begin{table}[htb]
\[
\begin{array}{|r||l|l|l|l|}
\hline
\vphantom{\Big|} n&\ \kwb0n &\ \kwb1n &\ \kwb2n &\ \kwb3n\\ \hline
0& (1,\ 0,\ 0,\ 0)& (0,\ 1,\ 0,\ 0)&
\ ( 0,\ 0,\ 1,\ 0)& (0,\ 0,\ 0,\ 1)\\ \hline
1& (0,\ 1,-1,\ 1)& (1,\ 0,-1,-1)&
(1,-1,\ 0,\ 1)& (1,\ 1,\ 1,\ 0)\\
2& (0,\ 1,-\omega,\ \cc\omega)&
(1,\ 0,-\omega,-\cc\omega)&
(1,-\omega,\ 0,\ \cc\omega)&
(1,\ \omega,\ \cc\omega,\ 0)\\
3& (0,\ 1,-\cc\omega,\ \omega)&
(1,\ 0,-\cc\omega,-\omega)&
(1,-\cc\omega,\ 0,\ \omega)&
(1,\ \cc\omega,\ \omega,\ 0)\\ \hline
4& (0,\ 1,-\omega,\ 1)&
(1,\ 0,-1,-\omega)&
(1,-\cc\omega,\ 0,\ \cc\omega)&
(1,\ \omega,\ 1,\ 0)\\
5& (0,\ 1,-\cc\omega,\ \cc\omega)&
(1,\ 0,-\omega,-1)&
(1,-1,\ 0,\ \omega)&
(1,\ \cc\omega,\ \cc\omega,\ 0)\\ 
6& (0,\ 1,-1,\ \omega)&
(1,\ 0,-\cc\omega,-\cc\omega)&
(1,-\omega,\ 0,\ 1)&
(1,\ 1,\ \omega,\ 0)\\ \hline
7& (0,\ 1,-\cc\omega,\ 1)&
(1,\ 0,-1,-\cc\omega)&
(1,-\omega,\ 0,\ \omega)&
(1,\ \cc\omega,\ 1,\ 0)\\
8& (0,\ 1,-1,\ \cc\omega)&
(1,\ 0,-\omega,-\omega)&
(1,-\cc\omega,\ 0,\ 1)&
(1,\ 1,\ \cc\omega,\ 0)\\
9& (0,\ 1,-\omega,\ \omega)&
(1,\ 0,-\cc\omega,-1)&
(1,-1,\ 0,\ \cc\omega)&
(1,\ \omega,\ \omega,\ 0) \\ \hline
 \end{array}
\]
\caption{Indexes for 40 states (up to normalization).}
\label{tab40rays}
\end{table}

Let us note that four triples of states  
$$\kwb00,\kwb10,\kwb20;\quad \kwb31,\kwb32,\kwb33;
\quad \kwb34,\kwb35,\kwb36;\quad \kwb37,\kwb38,\kwb39$$ 
are {\em mutually unbiased bases} (MUB) \cite{Woot89,low}
for 3D subspace of vectors with zero last coordinate.
Other states in the Table~\ref{tab40rays} are obtained
by cyclic shift of coordinates and appropriate change of signs
in agreement with \Eq{rays40a}.
Each state $\kwb{k}{n}$ is orthogonal with 12 states from 4 bases in
Table~\ref{tab40rays} and `unbiased' with other 27 states, {\em i.e.},
 $|\brkt{\wb{k}{n}}{\wb{l}{m}}|^2$ is either $0$ or $1/3$.

\subsection{Triflections}
\label{Sec:trif}

{\em Witting polytope} \Eq{points240} was described by H.~Coxeter 
\cite{CoxReg,Cox84} using four `unitary reflection'
matrices $R_k$ with property $R_k^3 = 1$
\begin{equation}
\begin{split}
R_1 = \begin{pmatrix}
\omega &0 &0 &0 \\
0 &1 &0 &0 \\
0 &0 &1 &0 \\
0 &0 &0 &1
\end{pmatrix}, &\quad 
R_2 = \frac{-\mi \cc{\omega}}{\sqrt{3}}
\begin{pmatrix}
\cc{\omega} &1 &1 &0 \\
1 &\cc{\omega} &1 &0 \\
1 &1 &\cc{\omega} &0 \\
0 &0 &0 &\mi \omega \sqrt{3}
\end{pmatrix},
\\
R_3 = \begin{pmatrix}
1 &0 &0 &0 \\
0 &1 &0 &0 \\
0 &0 &\omega &0 \\
0 &0 &0 &1
\end{pmatrix},&\quad 
R_4 = \frac{-\mi \cc{\omega}}{\sqrt{3}}
\begin{pmatrix}
\mi \omega \sqrt{3}&0 &0 &0 \\ 
0 &\cc{\omega} &-1 &1  \\
0 &-1 &\cc{\omega} &-1 \\
0 &1 &-1 &\cc{\omega}
\end{pmatrix}.
\label{Rmatr}
\end{split}
\end{equation}

Group of symmetries of Witting polytope with 155520 elements
is generated by matrices $R_k$ \cite{CoxReg,Cox84}.
Each complex ray \Eq{rays40} or state $\ket{\phi_n^k}$ corresponds to six vertexes of 
the polytope with different phase multipliers $\pm\omega^\mu$.

The matrices \Eq{Rmatr} are complex reflections \cite{CoxReg}
of order $k=3$ and for some ray $\ket{\phi}$ and root of unity $\zeta^k=1$ 
of arbitrary order 
\begin{equation}
R_\phi = \Id + (\zeta-1)\frac{\ket{\phi}\!\bra{\phi}}{\brkt{\phi}{\phi}}.
\label{CRefl}
\end{equation}
In \Eq{Rmatr} $\zeta=\omega$ and four rays
are $\kwb00, \kwb31, \kwb20, \kwb01$ respectively.

The determinants of all matrices \Eq{Rmatr} are $\omega$.
Let us introduce matrices  
\begin{equation}
r_k = \cc{\omega} R_k,\quad k = 1,\dots,4
\label{rmatr}
\end{equation}
with unit determinants, $(\cc{\omega})^4 \omega = 1$.
Such matrices generate group denoted further as $\Wit$
with 51840 elements.

\subsection{Symmetries of Witting configuration}
\label{Sec:sym}

Group of projective transformations of $\mathbb{CP}^3$ 
mentioned in \cite{ZP} is factor group $\Wt = \Wit/\{+\Id,-\Id\}$ 
with $51840/2=25920$ elements. 
It is isomorphic with finite
group $\mathtt{U_4(2)}$ of unitary $4\times 4$
matrices over Galois field $\mathbb{F}_{4}$ \cite[p. 26]{Atl}.

Formally, matrix $-\Id \in \Wit$ converts a ray or state into an equivalent one
and so, group $\Wt$ could be formally treated as a more proper way to represent symmetries 
of Witting configuration. 

Alternative description of Witting configuration with finite field $\mathbb{F}_4$
regarded as quadratic extension $\mathbb{F}_2$ can be found in \cite{CoxReg}.
With some abuse of notation four elements of $\mathbb{F}_4$ can be written
as $0$, $1$, $\omega$, $\omega^2=\cc\omega$, $\omega^3=1$ with addition and multiplication tables
\begin{equation}\label{GF4tabs}
\begin{array}{|c||c|c|c|c|}\hline
 +         & 0         & 1          & \omega    & \cc\omega\\ \hline\hline
 0         & 0         & 1          & \omega    & \cc\omega\\ \hline
 1         & 1         & 0          & \cc\omega & \omega \\ \hline
 \omega    & \omega    & \cc\omega  & 0         & 1 \\ \hline
 \cc\omega & \cc\omega & \omega     & 1         & 0   \\ \hline
\end{array}
\qquad
\begin{array}{|c||c|c|c|c|}\hline
\times    & 0 & 1         & \omega    & \cc\omega\\ \hline\hline
0         & 0 & 0         & 0         & 0      \\ \hline
1         & 0 & 1         & \omega    & \cc\omega \\ \hline
\omega    & 0 & \omega    & \cc\omega & 1        \\ \hline
\cc\omega & 0 & \cc\omega & 1         & \omega   \\ \hline
\end{array}.
\end{equation}
Here Galois conjugation is introduced as $\cc{a} = a^2$ and for the field of
the characteristic two $a+a=0$, $a=-a$. Thus, 40 points representing Witting configuration
in finite projective space $PG(3,2^2)$ can be obtained directly from Table~\ref{tab40rays}
by dropping all minus signs.

The construction of representation $\Wt \cong \mathtt{U_4(2)}$ is rather straightforward in such a model.
The $4 \times 4$ matrices with coefficients from $\mathbb{F}_{4}$ generating the group can be 
written using analogue of \Eq{CRefl} for appropriate rays to obtain matrices similar with
\Eq{Rmatr} or \Eq{rmatr} with extra $\cc{\omega}$ multiplier to set determinant
equivalent to unit in field $\mathbb{F}_4$ 
\begin{equation}
\begin{split}
	r_1^{(\mathbb{F}_4)} = \begin{pmatrix}
		1 &0 &0 &0 \\
		0 &\cc{\omega} &0 &0 \\
		0 &0 &\cc{\omega} &0 \\
		0 &0 &0 &\cc{\omega}
	\end{pmatrix}, &\quad 
	r_2^{(\mathbb{F}_4)} = 
	\begin{pmatrix}
		1      &\omega &\omega &0 \\
		\omega & 1     &\omega &0 \\
		\omega &\omega & 1 &0 \\
		0 &0 &0 & \cc{\omega}
	\end{pmatrix},
	\\
	r_3^{(\mathbb{F}_4)} = 
	 \begin{pmatrix}
		\cc{\omega} &0 &0 &0 \\
		0 &\cc{\omega} &0 &0 \\
		0 &0 & 1 &0 \\
		0 &0 &0 &\cc{\omega}
	\end{pmatrix},&\quad 
	r_4^{(\mathbb{F}_4)} = 
	\begin{pmatrix}
		\cc{\omega} &0 &0 &0 \\ 
		0 & 1     &\omega &\omega  \\
		0 &\omega & 1     &\omega \\
		0 &\omega &\omega & 1
	\end{pmatrix}.
	\label{rmatrGF4}
\end{split}
\end{equation}

Due to Galois conjugation Hermitian form $\sum_k u_k\cc{u}_k$ can be written as a cubic 
$\sum_k{u^3_k}$ in $\mathbb{F}_4$. Such representation provides some useful links with 
properties of cubic surfaces \cite{CoxReg}, but such relations are not discussed in
presented work.

\smallskip

The bigger group $\Wit$ can be more appropriate here due to existence of representation 
by unitary operators in 4D Hilbert space and some other reasons. The group
is isomorphic with finite group sometimes denoted as
$\mathtt{2.U_4(2)}$ \cite{Atl}.
The full group of symmetries of Witting polytope with
155520 elements  is denoted $\mathtt{3\mathsf{x}2.U_4(2)}$ in \cite[p.26]{Atl} 
and complex reflections \Eq{CRefl} of third order such as $R_k$ \Eq{Rmatr} 
are called {\em triflections}.

\subsection{Bases and contextuality}
\label{Sec:bas}

Witting configuration with 40 states from Table~\ref{tab40rays} can be arranged into 40 orthogonal 
tetrads (bases). They are listed in Table~\ref{tab40bas} in lexicographic 
order with respect to indexes $(l,n)$ of $\kwb{l}{n}$.

\begin{table}[htb]
	{\renewcommand{\arraystretch}{1.125}
	 \setlength{\arraycolsep}{0.1em}	
		\[
		\begin{array}{r@{:}lr@{:}lr@{:}l}
		1&\bigl\{\kwb00, \kwb01, \kwb02, \kwb03 \bigr\},&2& \bigl\{\kwb00, \kwb04, \kwb05, \kwb06 \bigr\}, &
		3&\bigl\{\kwb00, \kwb07, \kwb08, \kwb09 \bigr\},\\
		4&\lefteqn{\bigl\{\ul{ \kwb00, \kwb10, \kwb20, \kwb30 }\bigr\} 
			\ \equiv\ \bigl\{\ket0, \ket1, \ket2, \ket3\bigr\},}\\   
		5&\bigl\{\ul{ \kwb01, \kwb11, \kwb21, \kwb31 }\bigr\}, &6& \bigl\{\kwb01, \kwb16, \kwb24, \kwb35 \bigr\}, &
		7&\bigl\{\kwb01, \kwb18, \kwb27, \kwb39 \bigr\}, \\
		8&\bigl\{\kwb02, \kwb12, \kwb29, \kwb36 \bigr\}, &9& \bigl\{\kwb02, \kwb14, \kwb23, \kwb37 \bigr\}, &
		10&\bigl\{\ul{ \kwb02, \kwb19, \kwb26, \kwb32 }\bigr\}, \\
		11&\bigl\{\kwb03, \kwb13, \kwb25, \kwb38 \bigr\}, &12& \bigl\{\ul{ \kwb03, \kwb15, \kwb28, \kwb33 }\bigr\}, &
		13&\bigl\{\kwb03, \kwb17, \kwb22, \kwb34 \bigr\}, \\
		14&\bigl\{\kwb04, \kwb13, \kwb27, \kwb32 \bigr\}, &15& \bigl\{\kwb04, \kwb15, \kwb21, \kwb36 \bigr\}, & 
		16&\bigl\{\ul{ \kwb04, \kwb17, \kwb24, \kwb37 }\bigr\}, \\
		17&\bigl\{\kwb05, \kwb11, \kwb26, \kwb34 \bigr\}, &18& \bigl\{\ul{ \kwb05, \kwb16, \kwb29, \kwb38 }\bigr\}, & 
		19&\bigl\{\kwb05, \kwb18, \kwb23, \kwb33 \bigr\}, \\
		20&\bigl\{\ul{ \kwb06, \kwb12, \kwb22, \kwb39 }\bigr\}, &21& \bigl\{\kwb06, \kwb14, \kwb25, \kwb31 \bigr\}, &
		22&\bigl\{\kwb06, \kwb19, \kwb28, \kwb35 \bigr\}, \\
		23&\bigl\{\kwb07, \kwb12, \kwb24, \kwb33 \bigr\}, &24& \bigl\{\ul{ \kwb07, \kwb14, \kwb27, \kwb34 }\bigr\}, &
		25&\bigl\{\kwb07, \kwb19, \kwb21, \kwb38 \bigr\}, \\
		26&\bigl\{\ul{ \kwb08, \kwb13, \kwb23, \kwb35 }\bigr\}, &27& \bigl\{\kwb08, \kwb15, \kwb26, \kwb39 \bigr\}, &
		28&\bigl\{\kwb08, \kwb17, \kwb29, \kwb31 \bigr\}, \\
		29&\bigl\{\kwb09, \kwb11, \kwb28, \kwb37 \bigr\}, &30& \bigl\{\kwb09, \kwb16, \kwb22, \kwb32 \bigr\}, & 
		31&\bigl\{\ul{ \kwb09, \kwb18, \kwb25, \kwb36 }\bigr\}, \\ 
		32&\bigl\{\kwb10, \kwb11, \kwb12, \kwb13 \bigr\}, &33& \bigl\{\kwb10, \kwb14, \kwb15, \kwb16 \bigr\}, &
		34&\bigl\{\kwb10, \kwb17, \kwb18, \kwb19 \bigr\}, \\
		35&\bigl\{\kwb20, \kwb21, \kwb22, \kwb23 \bigr\}, &36& \bigl\{\kwb20, \kwb24, \kwb25, \kwb26 \bigr\}, &
		37&\bigl\{\kwb20, \kwb27, \kwb28, \kwb29 \bigr\}, \\
		38&\bigl\{\kwb30, \kwb31, \kwb32, \kwb33 \bigr\},\ &39& \bigl\{\kwb30, \kwb34, \kwb35, \kwb36 \bigr\},\ &
		40&\bigl\{\kwb30, \kwb37, \kwb38, \kwb39 \bigr\}. 
		\end{array}
		\] 
	}
	\caption{40 bases with states from Table~\ref{tab40rays}.}
	\label{tab40bas}
\end{table}

Each state belongs to four different bases including all 12 states orthogonal to given one.
The diagram with vertices corresponding to rays and edges connecting orthogonal rays 
is called Kochen-Specker graph (diagram) and the 40 bases represent {\em maximal 4-cliques} \cite{ZP}.
Such graphs are commonly used in discussions about contextuality in quantum
mechanics \cite{Pdod,ZP,MA,WA}.

Formally, any element of $\Wit$ up to permutations and multiplication on phases
can be associated with some basis represented by rows (or columns) of the 
unitary $4 \times 4$ matrix. There are $3^3=27$ diagonal matrices with coefficients $\omega^\nu$ 
and unit determinant. The exchange of basic vectors (with possible change of
directions) corresponds to subgroup of $\Wit$
with $2\cdot 4!=48$ elements generated by three matrices
$(R_k/R_{k+1})^2$, $k = 1,\ldots,3$
\begin{equation}\label{P_k}
\left(\begin{array}{rrrr}
\!-1 &0 &0 &0 \\ 
  0 &0 &-1&0  \\
  0 &-1&0 &0 \\
  0 &0 &0 &\ 1
 \end{array}\right)\!,\ %
\left(\begin{array}{rrrr}
  0 &-1&0 &0 \\ 
 \!-1 &0  &0 &0  \\
  0 &0 &-1& 0  \\
  0 &0 &0 &\ 1
\end{array}\right)\!,\ %
\left(\begin{array}{rrrr}
1 &0 &0 &0 \\ 
0 &0 &0 &-1  \\
0 &0 &-1&0  \\
0 &-1&0 &0
\end{array}\right)\!.
\end{equation}
Thus, all $40\cdot 27\cdot 48 = 51840$ elements of $\Wit$ can be represented in such a way.

\smallskip

Due to principles of quantum mechanics measurement of a state $\phi$ in any orthogonal basis 
should produce result corresponding to one of elements of this basis. Structure of 40 vectors
in Witting configuration makes impossible to suggest {\em non-contextual} deterministic classical
model of measurement corresponding to map of 40 states into set $\{0,1\}$ with requirement 
respecting a principle: {\em one and only one vector of any basis maps into $1$.} 

Such property follows from a correspondence with Penrose model with dodecahedra discussed earlier,
but here is useful to reintroduce some ideas directly from structure of Witting configuration.
Indeed, let us consider set of 10 bases underlined in Table~\ref{tab40bas}. The bases include all 40
rays of Witting configuration and thus, precisely 10 vectors should be mapped into 1. On the other 
hand, these 10 states may not include orthogonal vectors, because any such pair could be complemented
to full basis with at least two vectors mapped into 1.

Thus, necessary requirement could not be satisfied if it is not possible to find 10 nonorthogonal rays in
Witting configuration. This question also could be reformulated using {\em cliques}, but instead
of Kochen-Specker diagram mentioned above it should be used {\em dual graph} with edges
connecting {\em non-orthogonal} vertices. Straightforward test using GAP software \cite{GAP}
provides maximal size of cliques with non-orthogonal vectors is equal to {\em seven}.\hfill $\square$

In particular, two kinds of the non-orthogonal maximal cliques could be considered. There are
2880 different sets with seven non-orthogonal vectors mentioned above. Symmetry group $\Wit$ 
acts transitively on the cliques and any such set can be used to construct others, {\em e.g.}
\begin{equation}\label{cont7}
 \kwb00,\ \kwb11,\ \kwb14,\ \kwb17,\ \kwb32, \kwb35,\ \kwb38.
\end{equation}

Second kind with 90 maximal non-orthogonal cliques has only four states. They are also can be obtained
by application of elements of $\Wit$ to one of them, {\em e.g.},
\begin{equation}\label{cont4}
\kwb00,\ \kwb22,\ \kwb25,\ \kwb28.
\end{equation}
Any such tetrade belong to some plane in 4D Hilbert
space corresponding to a line in complex projective space $\mathbb{CP}^3$. Thus, 40 points
of Witting configuration lie by four on 90 projective lines \cite{ZP}.

\subsection{Measurements with single system}
\label{Sec:meas1}

Measurement of two entangled systems is discussed further in \Sec{meas2}.
For simplicity let us first consider two basic measurement schemes with single Witting
configuration adapted for possible implementation by quantum circuits. 
First one is measurement of a state in one of 40 bases. 

Let us consider an unitary matrix
$U_\Bas$ made up of four vectors in a basis $\Bas$. For convenience it may be chosen from 
symmetry group of Witting configuration $U_\Bas \in \Wit$, yet there are still
$51840/40 = 1296$ such matrices for each basis $\Bas$ due to phases and permutations \Eq{P_k}.
A smaller $\Wit_B \subset \Wit$  also can be used for the same purpose. The subgroup
may be generated by $r_1 r_3$ and $r_3 r_2 r_1 r_4$ together with $J_a$ defined
later in \Eq{Ja}. The subgroup has 1920 elements and less redundant for representation of bases, 
$1920/40=48$. Any basic state up to insignificant phase can be represented as
\begin{equation}\label{UBas}
 \ket{\phi_{(\Bas,k)}} = U_\Bas\ket{k},\quad k= 0,\dots,3.
\end{equation}
The quantum circuit with two qubits for such a scheme should use decomposition of 
matrix $U_\Bas^{-1}$ into set of elementary quantum gates to transform given state for final 
measurement in a computational basis.

Let us consider quantum system in one of 40 states $\kwb{l}{n}$ corresponding to Witting configuration.
A measurement without disturbance for given state could be performed using one of four bases
including of the state. Such a scheme is `contextual,' because construction of each basis used for 
measurement should include three extra states together with initial one. Such a method is
also has a problem with initial idea introduced by R. Penrose with possibility of
few steps, when tetrad of orthogonal states should not be chosen from very beginning.

Such a measurement scheme for some state $\ket\phi$ can be implemented using and auxiliary
qubit and unitary operator
\begin{equation}\label{Cmeas}
 C\!M_\phi = P_\phi \otimes X  + (\Id - P_\phi) \otimes \Id , \quad
 P_\phi = \ket{\phi}\!\bra{\phi}.
\end{equation}
Here operator $X$ denotes {\tt NOT} gate for an auxiliary qubit controlled by state of other two qubits
representing ququart.
The Toffoli gate is a particular case with $\kwb30=\ket{3}=\ket{1}\ket{1}$.  
Such a method is not `contextual,' because it does not suppose description of 
three extra states together with given $\ket{\phi}$. However contextuality can be also 
included in such approach if to use consequent application of different operators \Eq{Cmeas} 
with orthogonal states.

\section{Entanglement of Witting configurations}
\label{Sec:Ent}

\subsection{An extension of model with dodecahedra}
\label{Sec:ext}

Essential property of model discussed in \cite{Pdod,ZP,shadows} is consideration
of two entangled systems. 
In initial model with two spin-3/2 particles it was suggested to consider so-called `singlet state' \cite{Pdod}
corresponding to entangled pair with total spin zero \cite{shadows}
\begin{equation}\label{singlet}
 \ket{\Omega}=\ket{{\uparrow\uparrow\uparrow}}\ket{{\downarrow\downarrow\downarrow}}-
  \ket{{\downarrow\uparrow\uparrow}}\ket{{\uparrow\downarrow\downarrow}}+
  \ket{{\downarrow\downarrow\uparrow}}\ket{{\uparrow\uparrow\downarrow}}-
  \ket{{\downarrow\downarrow\downarrow}}\ket{{\uparrow\uparrow\uparrow}}.
\end{equation}

The state \Eq{singlet} is invariant with respect to transformation of two spin-3/2 particles due to
spatial rotations, {\em i.e.}, representation of SU(2) as some subgroup in SU(4). Due to such representation
{\em binary dodecahedral} (or binary icosahedral) group with 120 elements (corresponding to symmetries of dodecahedron
or icosahedron) can be mapped into transformations of Witting configuration discussed earlier in \cite{MA,WA}. 
The group is double cover of usual icosahedral group in SO(3) with 60 spatial rotations.

However such group describes directly only transformation of 20 states and more symmetries
can be used instead.
An entangled state of two systems can be described using some matrix $J$
\begin{equation}\label{EntJ}
 \ket{\Omega_J} = \mu_J \sum_{jk} J_{jk}\ket{j}\ket{k}, 
\end{equation}
where $\mu_J = \mathrm{Tr}(J\!J^*)^{-1/2}$ is a multiplier for normalization.
Let us apply the same unitary transformation $A$ to both systems
\begin{equation}\label{Ajk}
  \ket{j'}= \sum_{lj}A_{lj}\ket{j},\quad \ket{k'}=\sum_{mk}A_{mk}\ket{k}.
\end{equation}
In such a case \Eq{EntJ} is transformed as
\begin{equation}\label{EntJAA}
\ket{\Omega_J'} = \mu_J \sum_{jklm} J_{jk}A_{lj}A_{mk}\ket{j}\ket{k} 
\end{equation}
and $\ket{\Omega_J'}=\ket{\Omega_J}$ if
\begin{equation}\label{cAJJA}
 J = A J A^{T} \Longrightarrow \cc{A} J = J A,
\end{equation}
where $A^T$ is transposed matrix and $\cc{A}= (A^T)^{-1}$ for unitary $A$.

The simplest analogue of \Eq{singlet} is `antisymmetric' entangled state
\begin{equation}\label{entW}
\ket{\Omega_a}=\frac{1}{2}\bigl(\ket{0}\ket{3}-\ket{1}\ket{2}+\ket{2}\ket{1}-\ket{3}\ket{0}\bigr)
\end{equation}
with matrix
\begin{equation}\label{Ja}
 J_a = \begin{pmatrix}
  0 & 0 & 0 & 1\\
  0 & 0 & -1& 0\\
  0 & 1 & 0 & 0\\
  -1& 0 & 0 & 0
 \end{pmatrix}.
\end{equation}
Two alternative examples with matrices $J_1$ \Eq{J1} and $J_2$ \Eq{J2} are discussed later.

\smallskip

Similar condition \Eq{cAJJA} with slightly different matrix
\begin{equation}\label{Jpost}
J = \begin{pmatrix}
0 & 0 & 0 & 1\\
0 & 0 & 1& 0\\
0 & -1 & 0 & 0\\
-1& 0 & 0 & 0
\end{pmatrix}
\end{equation}
was also used in \cite{Post} for identification of
Spin(5) group with Sp(2). The matrix $J$
would correspond to entangled state
\begin{equation}\label{entWp}
 \frac{1}{2}\bigl(\ket{0}\ket{3}+\ket{1}\ket{2}-\ket{2}\ket{1}-\ket{3}\ket{0}\bigr)
\end{equation}
that can be converted into $\ket{\Omega_a}$ by swap of two basic states
$\ket{1}\leftrightarrow\ket{2}$. However, such a swap formally did not saves
Witting configuration with chosen positions of signs for some coordinates.

Anyway, subgroup of transformations respecting `antisymmetric' state $\ket{\Omega_a}$ is also
isomorphic with Spin(5) group described by $10$ real parameters. On the other hand, `symmetric'
state 
\begin{equation}\label{entWs}
\frac{1}{2}\bigl(\ket{0}\ket{0}+\ket{1}\ket{1}+\ket{2}\ket{2}+\ket{3}\ket{3}\bigr)
\end{equation} 
is invariant with respect to group SO(4) with only 6 real parameters.
It may justify consideration of models with antisymmetric entangled states 
despite of certain complication.

\smallskip

One subtlety is an analogy of `an opposite state' used in models with two dodecahedra for description
of joint measurement. Let us prove that for entangled state
\Eq{EntJ} with matrix such as  $J_a$ in \Eq{Ja} 
the `$J$-opposite' state 
can be expressed by {\em anti-unitary map}
\begin{equation}\label{Jopp}
  \ket{\psi_J} \simeq J \ket{\cc{\psi}}.
\end{equation}
An equivalence in \Eq{Jopp} is correctly defined up to unessential phase multiplier.
For basic states $\ket{k}$, $k=0,\dots,3$ the \Eq{Jopp} is valid for $J$ or $J_a$
up to unessential signs.
Let us consider some transformation $A_J$ satisfying \Eq{cAJJA} for given $J$
\[ \ket{\psi'} = A_J{\ket{\psi}}. \]
In such a case for $J$-opposite state \Eq{Jopp} 
\begin{equation}\label{AJopp}
A_J \ket{\psi_J} \simeq A_J J \ket{\cc{\psi}} = J \cc{A}_J\ket{\cc{\psi}}  
= J \ket{\cc{\psi'}} \simeq \ket{\psi'_J} 
\end{equation}
and so, $J$-opposite states $\ket{\psi}$ and $\ket{\psi_J}$ are transformed
by operator $A$ into $J$-opposite states $\ket{\psi'}$ and $\ket{\psi_J'}$.
The expression \Eq{AJopp} also illustrate that map \Eq{Jopp} should be
anti-unitary due to structure of \Eq{cAJJA}.

List of $J_a$-opposite states for Witting configuration with indexing from
Table~\ref{tab40rays} is shown in Table~\ref{tab40opp}.

\begin{table}[htb]
{\renewcommand{\arraystretch}{1.125}	
\[
 \begin{array}{|c|c|c|c|}
 \hline
 \kwb00 \leftrightarrow \kwb30 & \kwb10 \leftrightarrow \kwb20 &
  \kwb20 \leftrightarrow \kwb10 & \kwb30 \leftrightarrow \kwb00 \\ \hline
 \kwb01 \leftrightarrow \kwb31 & \kwb11 \leftrightarrow \kwb21 &
  \kwb21 \leftrightarrow \kwb11 & \kwb31 \leftrightarrow \kwb01 \\ \hline
 \kwb02 \leftrightarrow \kwb32 & \kwb12 \leftrightarrow \kwb22 &
  \kwb22 \leftrightarrow \kwb12 & \kwb32 \leftrightarrow \kwb02 \\ \hline
 \kwb03 \leftrightarrow \kwb33 & \kwb13 \leftrightarrow \kwb23 &
  \kwb23 \leftrightarrow \kwb13 & \kwb33 \leftrightarrow \kwb03 \\ \hline
 \kwb04 \leftrightarrow \kwb37 & \kwb14 \leftrightarrow \kwb27 &
  \kwb24 \leftrightarrow \kwb17 & \kwb34 \leftrightarrow \kwb07 \\ \hline
 \kwb05 \leftrightarrow \kwb38 & \kwb15 \leftrightarrow \kwb28 &
  \kwb25 \leftrightarrow \kwb18 & \kwb35 \leftrightarrow \kwb08 \\ \hline
 \kwb06 \leftrightarrow \kwb39 & \kwb16 \leftrightarrow \kwb29 &
  \kwb26 \leftrightarrow \kwb19 & \kwb36 \leftrightarrow \kwb09 \\ \hline
 \kwb07 \leftrightarrow \kwb34 & \kwb17 \leftrightarrow \kwb24 &
  \kwb27 \leftrightarrow \kwb14 & \kwb37 \leftrightarrow \kwb04 \\ \hline
 \kwb08 \leftrightarrow \kwb35 & \kwb18 \leftrightarrow \kwb25 &
  \kwb28 \leftrightarrow \kwb15 & \kwb38 \leftrightarrow \kwb05 \\ \hline
 \kwb09 \leftrightarrow \kwb36 & \kwb19 \leftrightarrow \kwb26 &
  \kwb29 \leftrightarrow \kwb16 & \kwb39 \leftrightarrow \kwb06 \\ \hline  
 \end{array}
\]
\caption{Pairs of `$J_a$-opposite' states.}
\label{tab40opp}
}	
\end{table}

\smallskip

Let us note, that \Eq{cAJJA} is true for any element of some group 
if it is satisfied for generators. An example is a subgroup of $\Wit$ 
with 720 elements generated by two matrices
\begin{equation}\label{Hgens}
 r_1 r_3 = \begin{pmatrix}
 \cc{\omega} & 0 & 0 & 0\\
 0 & \omega & 0& 0\\
 0 & 0 & \cc{\omega} & 0\\
 0& 0 & 0 & \omega
 \end{pmatrix}\!, \quad
 r^{}_2 r_4^{-1} = \frac{-\mi}{\sqrt{3}} 
 \begin{pmatrix}
 \omega & \cc{\omega} & \cc{\omega} & 0\\
 \cc{\omega} & 0 & -1 & -\omega\\
 \cc{\omega} & -1 & 0 & \omega \\
 0 & -\omega & \omega & -\cc{\omega}
 \end{pmatrix}.
\end{equation}
The group is denoted further as $\HWit$. 
It is isomorphic with finite group $\mathrm{SL}_2(\mathbb{F}_9)$ 
also known as double cover of $A_6$ (even permutations on six elements).
Let us point some analogy with mentioned earlier binary dodecahedral 
group as a double cover of $A_5$ (even permutations on five elements).

The condition \Eq{cAJJA} can be extended due to insignificance of
phase multipliers, $\ket{\Omega_J'}=\me^{\phi}\ket{\Omega_J}$.
The simplest case $\ket{\Omega_J'}=\pm\ket{\Omega_J}$ corresponds to
generalization of \Eq{cAJJA}
\begin{equation}\label{pmAJJA}
J = \pm A J A^{T} \Rightarrow \cc{A} J = \pm J A.
\end{equation}
For example, yet another group $\HZWit \cong \mathrm{SL}_2(\mathbb{F}_9)\rtimes C_2$ 
with 1440 elements can be constructed using two generators from
\Eq{Hgens} together with third generator $J_2$, see \Eq{J2}.
Such a matrix is satisfying \Eq{pmAJJA} {\em with minus sign}.

\smallskip

Let us note, that measurements of entangled state in any pair of bases obtained by
transformation $A_J$ from such $J_a$-invariant groups as $\HWit$ or $\HZWit$
should produce a pair of $J_a$-opposite states and so such bases should
contain two pairs of $J_a$-opposite states from Table~\ref{tab40opp}. 
It may be checked directly, that only 10 underlined bases in 
Table~\ref{tab40rays} have such properties.

\medskip

The more general example can be obtained, if for given $J$ to apply arbitrary unitary
transformation $A$ to first system and for second one to find $B$
saving state $\ket{\Omega_J}$ \Eq{EntJ} invariant.
Thus, instead of \Eq{EntJAA} an expression for two different
transformation should be written
\begin{equation}\label{EntJAB}
\ket{\Omega_J'} = \mu_J \sum_{jklm} J_{jk}A_{lj}B_{mk}\ket{j}\ket{k}=\ket{\Omega_J} 
\end{equation}
and equations with $B$ similar with \Eq{cAJJA} can be simply obtained. 
For invertible matrices $J$ it can be also written
\begin{equation}\label{cAJJB}
J = A J B^{T} \Longrightarrow \cc{B} J = J A,\quad B = \cc{J A J^{-1}}
\end{equation}
and for real matrices $J=\cc{J}$ the expressions can be simplified
\begin{equation}\label{BJJcA}
 B J = J \cc{A},\quad B = J \cc{A} J^{-1}.
\end{equation}
The \Eq{Jopp} for $J$-opposite state is not changed in such a case, but
now pair of states are transformed by operators $A_J$ and $B_J$ respectively,
{\em cf.\/} \Eq{AJopp}
\begin{equation}\label{ABopp}
\ket{\psi'} = A_J{\ket{\psi}},\quad
\ket{\psi'_J} \simeq J \ket{\cc{\psi'}} = J \cc{A}_J\ket{\cc{\psi}} =
B_J J \ket{\cc{\psi}} \simeq B_J \ket{\psi_J}. 
\end{equation}

For 30 bases between 40 a pair representing tetrad with $J_a$-opposite states
does not coincide with initial base. Thus a different matrices
$A$ an $B$ should be used for transformations of such bases.
For Table~\ref{tab40bas} the indexes for all 40 pairs of bases are listed in 
Table~\ref{tabOppBas}.

\begin{table}[htb]
\[
\begin{array}{cccccccc}
( 1 : 38 ), &( 2 : 40 ), &( 3 : 39 ), &(\ul{ 4 : 4 }),  &(\ul{ 5 : 5 }),\\
( 6 : 28 ), &( 7 : 21 ), &( 8 : 30 ), &( 9 : 14 ), &(\ul{ 10 : 10 }),\\
( 11 : 19 ), &(\ul{ 12 : 12 }), &( 13 : 23 ), &( 14 : 9 ), &( 15 : 29 ),\\
(\ul{ 16 : 16 }), &( 17 : 25 ), &(\ul{ 18 : 18 }), &( 19 : 11 ), &(\ul{ 20 : 20 }),\\
( 21 : 7 ), &( 22 : 27 ), &( 23 : 13 ), &(\ul{ 24 : 24 }), &( 25 : 17 ),\\
(\ul{ 26 : 26 }), &( 27 : 22 ), &( 28 : 6 ), &( 29 : 15 ), &( 30 : 8 ), \\
(\ul{ 31 : 31 }), &( 32 : 35 ), &( 33 : 37 ),&( 34 : 36 ), &( 35 : 32 ),\\
( 36 : 34 ), &( 37 : 33 ), &( 38 : 1 ), &( 39 : 3 ), &( 40 : 2 ).
\end{array}
\] 
\caption{Indexes of bases with `$J_a$-opposite' states in Table~\ref{tab40bas}.}
\label{tabOppBas}
\end{table}

\subsection{Measurements of two entangled systems}
\label{Sec:meas2}

Schemes of measurement adapted for single system of states from Witting
configuration were outlined in \Sec{meas1}. 
A measurement of two such systems with the same setup is appropriate for 10 bases
underlined in Table~\ref{tab40bas} corresponding to symmetries from subgroups 
$\HWit$ or $\HZWit$. The bases contain all 40 states and could be used
for simple demonstration of basic property of measurements with
entangled state $\ket{\Omega_a}$ \Eq{entW} : if one system
after measurement is found in some state $\kwb{l}{n}$ the second one
has $J_a$-opposite state from Table~\ref{tab40opp}.

However, 10 bases are not enough for formulation of some problems related with contextuality.
More general setup suggests possibility to select some state $\kwb{l}{n}$ for first system
using scheme with $C\!M_\phi$ operator \Eq{Cmeas} together with $J_a$-opposite state for second system.
There are four bases (between 40) containing any state $\kwb{l}{n}$, but in subset with 10 bases mentioned 
above there is only one basis with any given state. 

Only consideration of all 40 bases let us choose for any state $\kwb{l}{n}$ one between 
four bases for first system together with corresponding basis with $J_a$-opposite states for second 
one, see Table~\ref{tabOppBas}. Such approach is more close analogue of `multi-step' scheme of measurements 
suggested for the model with two dodecahedra in \cite{Pdod,ZP}.

\subsection{Other entangled states}
\label{Sec:alt}

The entangled state  $\ket{\Omega_a}$ \Eq{entW} is not an only possible
analogue of state \Eq{singlet} used in a model with two dodecahedra \cite{Pdod,ZP}. 
Let us consider some symmetry $S \in \Wit$ of Witting configuration
\begin{equation}\label{Spsi}
\ket{\psi} \mapsto \ket{\psi^S} = S \ket{\psi}.
\end{equation}
For a subgroup such as $\HZWit$ isomorphic subgroups $\HZWit^S$ can be constructed
using conjugations of all elements
\begin{equation}\label{conjSH}
 H \mapsto H^S = S H S^{-1}
\end{equation}
and there are $[\Wit:\HZWit] = 51840/1440 = 36$ such subgroups with $S$ from different cosets
$S\HZWit$. For $H \in \HZWit$ and \Eq{cAJJA} can be rewritten as
\begin{equation}\label{cHJHJ}
  \cc{H} = J^{-1} H J 
\end{equation}
and due to \Eq{conjSH}
$$ \cc{S}^{-1} \cc{H}^S \cc{S} = J^{-1} S^{-1} H^S S J$$
and thus instead of $J$ for group $\HZWit^S$ should be used matrix
\begin{equation}\label{JST}
 J^S = S J \cc{S}^{-1} = S J S^T.
\end{equation} 

Let us consider simplest cases with $J^S$ are matrices with 
coefficients 0 and $\pm1$. Together with initial example with $S=\Id$ and $J_a$ \Eq{Ja} only
two cases between 35 satisfy such conditions.
\begin{equation}\label{J1}
J_1 = \begin{pmatrix}
\ 0 &\ 0 &\ 1 &\ 0\\
\ 0 &\ 0 &\ 0 &\ 1\\
\!-1  &\ 0 &\ 0 &\ 0\\
\ 0 & \!-1 &\ 0 &\ 0
\end{pmatrix}
\end{equation}
and
\begin{equation}\label{J2}
J_2 = \begin{pmatrix}
\ 0 &\ 1 &\ 0 &\ 0\\
\!-1 &\ 0 &\ 0 &\ 0\\
\ 0 &\ 0 &\ 0 &\!-1\\
\ 0 &\ 0 &\ 1 &\ 0
\end{pmatrix}.
\end{equation}
Entangled states for such matrices are
\begin{equation}\label{entW1}
\ket{\Omega_1}=\frac{1}{2}\bigl(\ket{0}\ket{2}-\ket{2}\ket{0}+\ket{1}\ket{3}-\ket{3}\ket{1}\bigr)
\end{equation}
and
\begin{equation}\label{entW2}
\ket{\Omega_2}=\frac{1}{2}\bigl(\ket{0}\ket{1}-\ket{1}\ket{0}+\ket{3}\ket{2}-\ket{2}\ket{3}\bigr).
\end{equation}

As transformations $S$ used for construction of $\HZWit^S$ can be chosen
\begin{equation}\label{S12}
S_1 = \begin{pmatrix}
0 & 1 & 0 &\ 0\\
1 & 0 & 0 &\ 0\\
0 & 0 & 1 &\ 0\\
0 & 0 & 0 &\!-1
\end{pmatrix},
\qquad
S_2 = \begin{pmatrix}
1 & 0 & 0 & 0\\
0 & 0 & 0 & -1\\
0 & 0 & -1 & 0\\
0 & -1 & 0 & 0
\end{pmatrix}.
\end{equation}

\begin{table}[htbp]
	{\renewcommand{\arraystretch}{1.125}	
		\[
		\begin{array}{|c|c|c|c|}
		\hline
		\kwb00 \leftrightarrow \kwb20 & \kwb10 \leftrightarrow \kwb30 &
	 	 \kwb20 \leftrightarrow \kwb00 & \kwb30 \leftrightarrow \kwb10 \\ \hline
		\kwb01 \leftrightarrow \kwb21 & \kwb11 \leftrightarrow \kwb31 &
		 \kwb21 \leftrightarrow \kwb01 & \kwb31 \leftrightarrow \kwb11 \\ \hline
		\kwb02 \leftrightarrow \kwb23 & \kwb12 \leftrightarrow \kwb33 &
		 \kwb22 \leftrightarrow \kwb03 & \kwb32 \leftrightarrow \kwb13 \\ \hline
		\kwb03 \leftrightarrow \kwb22 & \kwb13 \leftrightarrow \kwb32 &
		 \kwb23 \leftrightarrow \kwb02 & \kwb33 \leftrightarrow \kwb12 \\ \hline
		\kwb04 \leftrightarrow \kwb27 & \kwb14 \leftrightarrow \kwb37 &
		 \kwb24 \leftrightarrow \kwb07 & \kwb34 \leftrightarrow \kwb17 \\ \hline
		\kwb05 \leftrightarrow \kwb29 & \kwb15 \leftrightarrow \kwb39 &
		 \kwb25 \leftrightarrow \kwb09 & \kwb35 \leftrightarrow \kwb19 \\ \hline
		\kwb06 \leftrightarrow \kwb28 & \kwb16 \leftrightarrow \kwb38 &
		 \kwb26 \leftrightarrow \kwb08 & \kwb36 \leftrightarrow \kwb18 \\ \hline
		\kwb07 \leftrightarrow \kwb24 & \kwb17 \leftrightarrow \kwb34 &
		 \kwb27 \leftrightarrow \kwb04 & \kwb37 \leftrightarrow \kwb14 \\ \hline
		\kwb08 \leftrightarrow \kwb26 & \kwb18 \leftrightarrow \kwb36 &
		 \kwb28 \leftrightarrow \kwb06 & \kwb38 \leftrightarrow \kwb16 \\ \hline
		\kwb09 \leftrightarrow \kwb25 & \kwb19 \leftrightarrow \kwb35 &
		 \kwb29 \leftrightarrow \kwb05 & \kwb39 \leftrightarrow \kwb15 \\ \hline  
		\end{array}
		\]
		\caption{Pairs of `$J_1$-opposite' states.}
		\label{tab40J1opp}
	}	
\end{table}

\begin{table}[htbp]
	{\renewcommand{\arraystretch}{1.125}	
		\[
		\begin{array}{|c|c|c|c|}
		\hline
		\kwb00 \leftrightarrow \kwb10 & \kwb10 \leftrightarrow \kwb00 &
		\kwb20 \leftrightarrow \kwb30 & \kwb30 \leftrightarrow \kwb20 \\ \hline
		\kwb01 \leftrightarrow \kwb11 & \kwb11 \leftrightarrow \kwb01 &
		\kwb21 \leftrightarrow \kwb31 & \kwb31 \leftrightarrow \kwb21 \\ \hline
		\kwb02 \leftrightarrow \kwb12 & \kwb12 \leftrightarrow \kwb02 &
		\kwb22 \leftrightarrow \kwb32 & \kwb32 \leftrightarrow \kwb22 \\ \hline
		\kwb03 \leftrightarrow \kwb13 & \kwb13 \leftrightarrow \kwb03 &
		\kwb23 \leftrightarrow \kwb33 & \kwb33 \leftrightarrow \kwb23 \\ \hline
		\kwb04 \leftrightarrow \kwb17 & \kwb14 \leftrightarrow \kwb07 &
		\kwb24 \leftrightarrow \kwb37 & \kwb34 \leftrightarrow \kwb27 \\ \hline
		\kwb05 \leftrightarrow \kwb18 & \kwb15 \leftrightarrow \kwb08 &
		\kwb25 \leftrightarrow \kwb38 & \kwb35 \leftrightarrow \kwb28 \\ \hline
		\kwb06 \leftrightarrow \kwb19 & \kwb16 \leftrightarrow \kwb09 &
		\kwb26 \leftrightarrow \kwb39 & \kwb36 \leftrightarrow \kwb29 \\ \hline
		\kwb07 \leftrightarrow \kwb14 & \kwb17 \leftrightarrow \kwb04 &
		\kwb27 \leftrightarrow \kwb34 & \kwb37 \leftrightarrow \kwb24 \\ \hline
		\kwb08 \leftrightarrow \kwb15 & \kwb18 \leftrightarrow \kwb05 &
		\kwb28 \leftrightarrow \kwb35 & \kwb38 \leftrightarrow \kwb25 \\ \hline
		\kwb09 \leftrightarrow \kwb16 & \kwb19 \leftrightarrow \kwb06 &
		\kwb29 \leftrightarrow \kwb36 & \kwb39 \leftrightarrow \kwb26 \\ \hline  
		\end{array}
		\]
		\caption{Pairs of `$J_2$-opposite' states.}
		\label{tab40J2opp}
	}	
\end{table}

List of $J_1$- and $J_2-$opposite states for Witting configuration with indexing from
Table~\ref{tab40rays} are shown in Table~\ref{tab40J1opp} and Table~\ref{tab40J2opp} 
respectively. Analogues of Table~\ref{tabOppBas} with pairs of bases written earlier for $J_a$ 
are Table~\ref{tabJ1OppBas} for $J_1$ and Table~\ref{tabJ2OppBas} for $J_2$.
Schemes of measurements discussed in \Sec{meas2} can be extended accordingly for
entangled states $\ket{\Omega_1}$ \Eq{entW1} and  $\ket{\Omega_2}$ \Eq{entW2}.

\begin{table}[htb]
	\[
	\begin{array}{cccccccc}
	( 1 : 35 ), &( 2 : 37 ), &( 3 : 36 ), &(\ul{ 4 : 4 }),  &(\ul{ 5 : 5 }),\\
	( 6 : 25 ), &( 7 : 15 ), &( 8 : 19 ), &(\ul{ 9 : 9 }), &(10 : 26 ),\\
	( 11 : 30 ), &( 12 : 20 ), &(\ul{ 13 : 13 }), &(\ul{ 14 : 14 }), &( 15 : 7 ),\\
	( 16 : 24 ), &( 17 : 28 ), &(\ul{ 18 : 18 }), &( 19 : 8 ), &( 20 : 12 ),\\
	( 21 : 29 ), &(\ul{ 22 : 22 }), &(\ul{ 23 : 23 }), &( 24 : 16 ), &( 25 : 6 ),\\
	( 26 : 10 ), &(\ul{ 27 : 27 }), &( 28 : 17 ), &( 29 : 21 ), &( 30 : 11 ), \\
	(\ul{ 31 : 31 }), &( 32 : 38 ), &( 33 : 40 ),&( 34 : 39 ), &( 35 : 1 ),\\
	( 36 : 3 ), &( 37 : 2 ), &( 38 : 32 ), &( 39 : 34 ), &( 40 : 33 ).
	\end{array}
	\] 
	\caption{Indexes of bases with `$J_1$-opposite' states in Table~\ref{tab40bas}.}
	\label{tabJ1OppBas}
\end{table}

\begin{table}[htb]
	\[
	\begin{array}{cccccccc}
	( 1 : 32 ), &( 2 : 34 ), &( 3 : 33 ), &(\ul{ 4 : 4 }),  &(\ul{ 5 : 5 }),\\
	( 6 : 29 ), &( 7 : 17 ), &(\ul{ 8 : 8 }), &( 9 : 23 ), &( 10 : 20 ),\\
	(\ul{ 11 : 11 }), &( 12 : 26 ), &( 13 : 14 ), &( 14 : 13 ), &( 15 : 28 ),\\
	(\ul{ 16 : 16 }), &( 17 : 7 ), &( 18 : 31 ), &(\ul{ 19 : 19 }), &( 20 : 10 ),\\
	( 21 : 25 ), &(\ul{ 22 : 22 }), &( 23 : 9 ), &(\ul{ 24 : 24 }), &( 25 : 21 ),\\
	( 26 : 12 ), &(\ul{ 27 : 27 }), &( 28 : 15 ), &( 29 : 6 ), &(\ul{ 30 : 30 } ), \\
	( 31 : 18 ), &( 32 : 1 ), &( 33 : 3 ),&( 34 : 2 ), &( 35 : 38 ),\\
	( 36 : 40 ), &( 37 : 39 ), &( 38 : 35 ), &( 39 : 37 ), &( 40 : 36 ).
	\end{array}
	\] 
	\caption{Indexes of bases with `$J_2$-opposite' states in Table~\ref{tab40bas}.}
	\label{tabJ2OppBas}
\end{table}

\newpage

\section{Conclusion}

Entanglement of two quantum systems described by Witting configurations with 40 states in 4D Hilbert space
is discussed in presented work. The model is originated from system of two entangled particles
with spin-3/2 suggested by Roger Penrose for formulation of analogue of Bell theorem \cite{Pdod,ZP,shadows}.
The initial model is known also as `Penrose dodecahedron,' because of a model with measurements 
is associated with vertices of two regular dodecahedra.

Here the description of measurements briefly outlined in \Sec{meas1} and \Sec{meas2} is reformulated
for more convenient implementation with quantum circuit model.
The structure of 40 rays in Witting configuration looks much more complicated in comparison with 
initial set with 20 rays. It was mentioned already in \cite{ZP} that such configuration has full 
symmetry $\Wt$ group with 25920 transformations. Such group can be represented as discrete subgroup in
group of unitary projective transformations of 4D complex space or as finite group $\mathtt{U_4(2)}$ 
\cite[p. 26]{Atl}. More convenient choice is double cover $\Wit$ with implementation by 51840 unitary 
$4 \times 4$ matrices discussed in this paper. The $\Wt$ can be naturally considered as factor group 
of $\Wit$ by the center with two matrices $\pm\Id$ representing trivial transformation of
projective space.

It can be mentioned for comparison, what group of symmetries of dodecahedron has only 60 transformations 
and double cover can be represented by 120 unitary $2 \times 2$ matrices for particle with spin-1/2
and representation with $4 \times 4$ unitary matrices used by R. Penrose correspond to spin-3/2 particle.
Initial model based on geometry of dodecahedron uses only symmetries induced by rotations in 3D space.
Approach suggested in presented work is more general with consideration of full group of symmetries
of 4D Hilbert space U$(4)$ and discrete subgroup $\Wit\cong\mathtt{2.U_4(2)}$ of symmetries of Witting 
configuration.

\end{document}